# Realistic wave-optics simulation of X-ray dark-field imaging at a human scale (Running title: wave-optics XDFI simulation)

Yongjin Sung,[a,*] Brandon Nelson[b], Rajiv Gupta[c]

[a]University of Wisconsin-Milwaukee, College of Engineering & Applied Science, 3200 North Cramer Street, Milwaukee, WI 53211, USA

[b]U.S. Food and Drug Administration, Center for Devices and Radiological Health, Office of Science and Engineering Labs, Division of Imaging, Diagnostics, and Software Reliability, 10903 New Hampshire Avenue, Silver Spring, MD 20993, USA

[c]Massachusetts General Hospital, Department of Radiology, 55 Fruit Street, Boston, MA 02114, USA

*Correspondence e-mail: ysung4@uwm.edu

## Abstract

Background: X-ray dark-field imaging (XDFI) has been explored to provide superior performance over the conventional X-ray imaging for the diagnosis of many pathologic conditions. A simulation tool to reliably predict clinical XDFI images at a human scale, however, is currently missing.

Purpose: In this paper, we demonstrate XDFI simulation at a human scale for the first time to the best of our knowledge. Using the developed simulation tool, we demonstrate the strengths and limitations of XDFI for the diagnosis of emphysema, fibrosis, atelectasis, edema, and pneumonia.

Methods: We augment the XCAT phantom with Voronoi grids to simulate alveolar substructure, responsible for the dark-field signal from lungs, assign material properties to each tissue type, and simulate X-ray wave propagation through the augmented XCAT phantom using a multi-layer wave-optics propagation. Altering the density and thickness of the Voronoi grids as well as the material properties, we simulate XDFI images of normal and diseased lungs.

Results: Our simulation framework can generate realistic XDFI images of a human chest with normal or diseased lungs. The simulation confirms that the normal, emphysematous, and fibrotic lungs show clearly distinct dark-field signals. It also shows that alveolar fluid accumulation in pneumonia, wall thickening in interstitial edema, and deflation in atelectasis result in a similar reduction in dark-field signal.

Conclusions: It is feasible to augment XCAT with pulmonary substructure and generate realistic XDFI images using multi-layer wave optics. By providing the most realistic XDFI images of lung pathologies, the developed simulation framework will enable in-silico clinical trials and the optimization of both hardware and software for XDFI.

**Introduction**

In conventional X-ray imaging, the image contrast arises from the attenuation of X-rays due to photoelectric absorption and Compton scattering. X-rays are weakly absorbed by low-atomic-number (low-Z) materials such as soft tissues. Thus, conventional X-ray imaging is unable to distinguish them despite providing excellent contrast for high-Z materials such as bones and metallic implants. Alternative approaches utilizing complementary contrast mechanisms to improve contrast in soft tissues have been explored for several decades. For example, X-ray phase-contrast imaging (XPCI) exploits X-ray refraction caused by local variations in the electron density[1]. It is well established that XPCI has significantly higher soft-tissue contrast than traditional attenuation-based X-ray imaging[2]. X-ray dark-field imaging (XDFI) exploits the scramble of coherent wavefronts due to sub-structure on the micrometer scale[3,4]. XDFI has been shown to provide superior performance over the conventional X-ray imaging for the diagnosis of many pathologic conditions such as pulmonary fibrosis and other pulmonary diseases[5–9]. Early methods for XPCI and XDFI required a highly coherent X-ray source such as synchrotron[10–14]. Consequently, their incorporation into routine medical imaging has been very difficult. Recent developments of Talbot-Lau interferometry have lifted this serious restriction[15], and it has become possible to build a tabletop XPCI/XDFI system using a conventional X-ray source[16,17]. Although early XPCI/XDFI systems were only suitable for small *ex-vivo* specimens[18] and small animals[19], considerable progress has been made to scale up the size of these systems for human imaging[20,21]. One of the key technical limitations in this endeavor is the lack of a reliable method for predicting the performance of these systems at a human scale.

A typical X-ray simulation framework consists of two parts: a numerical phantom that provides a detailed description of human anatomy and a forward model that implements the image formation process. Simulating XPCI/XDFI at human scale poses challenges for both parts of the simulation framework. For the simulation of medical X-ray imaging, voxelated numerical phantoms are typically used with the voxel size about the same as the detector pitch. For XPCI/XDFI simulation, the boundaries between the coarse voxels can generate strong diffraction artefact, which is hard to separate from a true XPCI/XDFI signal. In contrast, the 4D extended cardiac-torso (XCAT) phantom represents various organs using non-uniform rational B-spline (NURBS) surfaces[22,23], which can be used to generate a voxelated phantom at arbitrarily high resolution.

The forward model describes the interaction of incident X-rays with various anatomic structures in the numerical phantom as well as X-ray optical components. For XPCI simulation, ray tracing[24] or Monte Carlo methods[25], which are commonly used in conventional attenuation-based X-ray simulation, have been extended to calculate the X-ray refraction. Wave optics simulation can best model the wave phenomena such as coherent scattering and diffraction of X-rays; however, it usually requires large computational resources. We previously demonstrated a fast, yet accurate wave-optics simulation of XPCI using a forward model built upon the first-order Rytov approximation[26–28], which is valid when the refractive index difference at the boundary is small, and thus can be safely used in the XPCI simulation[29]. The so-called full-wave Rytov (FWR) method allows for fast yet accurate wave-optics simulation of XPCI using NURBS-defined models as well as geometric primitives and a voxelated phantom.

The first-order approximation, however, ignores multiple scattering within an imaged specimen, which can be problematic for XDFI simulation. This is because the image contrast of XDFI originates from the multiple scattering of X-rays by a microstructure such as the alveoli of lungs. Thus, for XDFI simulation, a multi-layer wave propagation method has been dominantly used[30–32], which can handle multiple scattering. In this method, a numerical phantom is divided into an axial stack of slices, and a wave propagation is performed across the axial stack consecutively. For the numerical propagation through each layer, the beam-propagation method (BPM) has been dominantly used, which uses the angular-spectrum scalar wave theory together with the projection approximation. With the projection approximation, the axial variation of the refractive index is ignored within the layer's volume. Although the angular-spectrum scalar wave theory can be implemented very efficiently using the Fourier transform[33], the grid size for the calculations must be very small to capture microstructure. This requirement significantly increases the computational cost. Thus, the XDFI simulation using the BPM has been mostly applied to a small phantom of relatively simple geometry such as a box filled with spheres[30,31].

For the numerical phantom, we generate a high-resolution voxelated phantom from the NURBS-defined XCAT model. The phantom is augmented by adding small air voids to the lung region in each slice, which generates the ultra-small angle X-ray scattering by alveoli, and thus the dark-field signal for lungs. For the forward model, we use a multi-layer wave propagation method as described in Methods.

## Methods

**Extended cardiac-torso (XCAT) phantom to simulate lung microstructure**

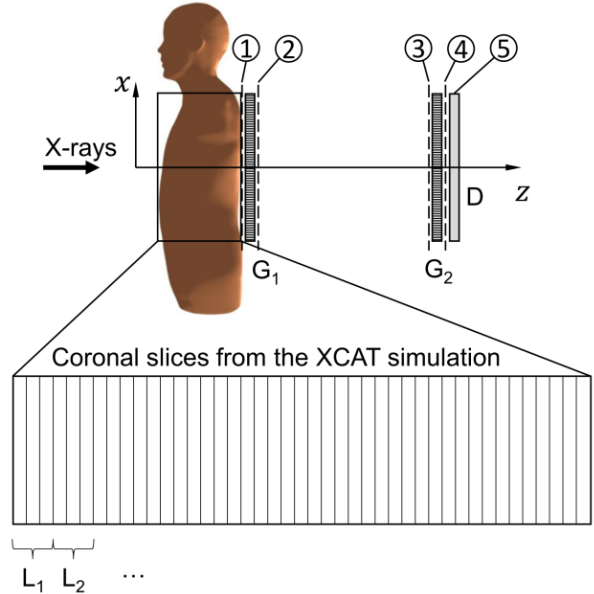

**Figure 1. Schematic diagram of the simulation geometry used in this study. G1, G2: gratings; and D: detector. The numbers 1 through 5 in the circles refer to the planes where Eqs. (6) through (9) were calculated.**

Figure 1 shows the simulation geometry used in this study. In a multi-layer wave-optics simulation, the numerical phantom is divided into an axial stack of layers, and the X-ray wave propagates through the layers in a sequential manner. For the numerical phantom, we use the XCAT phantom, which represents various organs with about 2800 NURBS surfaces[22,23], and thus can accurately model the complex human anatomy such as the lungs, skeleton, muscles, bronchial tree, and vasculature[34]. Due to its realism and computational efficiency, the XCAT phantom has been widely adopted to simulate various imaging modalities. Using the XCAT (Duke University, ver. 2), we produced 2505 coronal slices of 100 µm intervals. Each slice contained 4096×4096 pixels of 100 µm with the pixel value representing the organ identification number (ID) stored as a 16-bit integer.

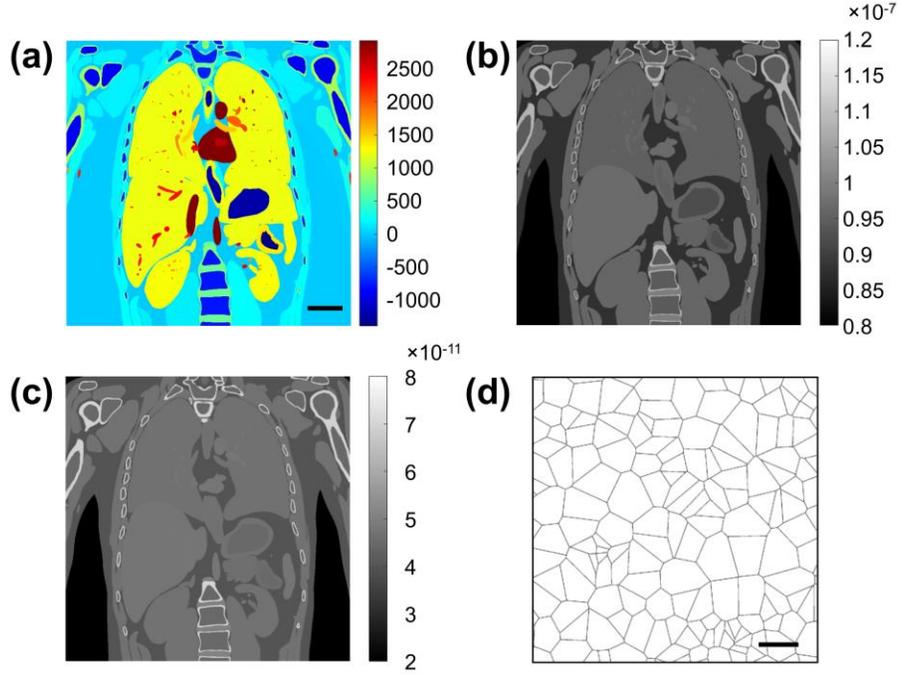

**Figure 2. Augmentation of the XCAT phantom for the XDFI simulation.** (a) shows an example slice of the XCAT, 4096×4096 pixels with the pixel size of 100 μm. (b) and (c) show the real and imaginary parts of the refractive index after assigning the material properties. (d) shows an example of the lung microstructure, which is modeled using Voronoi grids. The scale bar (5 cm) in (a) also applies to (b) and (c). Scale bar in (d): 1 mm.

Figure 2(a) shows an example coronal slice produced by the XCAT. Based on the organ ID, we have assigned the complex refractive index value $n(x, y, z)$ to each organ for the X-ray energy (i.e., wavelength) used for the simulation.

The complex refractive index, an input to the XPCI/XDFI simulation, is typically written in the form $n(x, y, z) = 1 - \delta(x, y, z) + i\beta(x, y, z)$.

$$\delta = N_{el} r_e \lambda^2 / 2\pi, \qquad [1]$$

$$\beta = \lambda \mu(\lambda) / 4\pi, \qquad [2]$$

where $\lambda$ is the wavelength of X-ray, $r_e$ is the classical radius of electron (2.818×10$^{-15}$ m). The electron density $N_{el}$ and the linear attenuation coefficient $\mu(\lambda)$ were obtained from the data for adult human tissues[35]. Figures 2(b) and 2(c) show $\delta$ and $\beta$, respectively, assigned to the coronal slice shown in Figure 2(a), for the X-ray energy of 50 keV.

The alveoli of human lungs have the mean diameter of 200 µm, and the lung parenchyma occupies about 92% of the total lung volume[36]. To model the microstructure of lungs, Taphorn et al. used 2D Voronoi grids[37], which we adopt here. To incorporate the microstructure into the XCAT phantom, we upsampled each XCAT slice by a factor of 80, which rendered the pixel resolution to be 1.25 µm, 1/8 of the grating period. Next, we replaced the regions identified as lungs with 2D Voronoi grids, then dilated the individual edges by the number of pixels corresponding to the wall thickness. Figure 2(d) shows an example of 2D Voronoi grids that we incorporated into the XCAT phantom. We assigned the material properties for lung tissue to the alveolus wall (i.e., Voronoi boundaries) and air to the inner region of each Voronoi cell (Table II). The alveolar density varies with the height (i.e., the distance from the diaphragm) in the lungs. McDonough et al. provided the alveolar density of $y = -2.5015x + 34.983$, where $y$ equals the number of alveoli and $x$ is the lung height (range, 1-5; 1: lung apex, 5: lung base); the alveolar density is 31.6 ± 3.4 alveoli/mm³ in the lung apex and 21.2 ± 1.6 alveoli/mm³ in the lung base. We have adopted the formula to assign 37 different values of alveolar density to the lungs depending on the height, so that the average alveolar density is 31.6 alveoli/mm³ in the upper one-fifth region and 21.2 alveoli/mm³ in the lower one-fifth region.

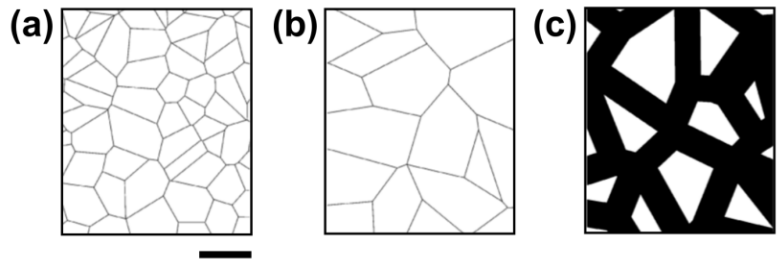

**Figure 3. Examples of the microstructures incorporated into the XCAT phantom are shown for normal (a), emphysematous (b), and fibrotic (c) lungs. The scale bar (1 mm) in (a) is also applied to (b) and (c).**

Figure 3 shows examples of the Voronoi grids for normal, emphysematous, and fibrotic lungs. For emphysematous lungs, the alveolar density was reduced to 80%, 50%, and 20% to simulate mild, moderate, and severe degree of emphysema, respectively, which resulted in enlarged alveoli correspondingly. For fibrotic lungs, the alveolar density was reduced to 1/4, and the wall thickness increased by 40 times. Further, we simulated atelectasis, edema, and pneumonia by augmenting the left lung of the XCAT phantom. For atelectatic lungs, the volume of the left lung was reduced by 10% along all the three dimensions, while maintaining the total number of alveoli. The lung was shrunk by modifying the control points of the NURBS model, from which the XCAT slices are generated.

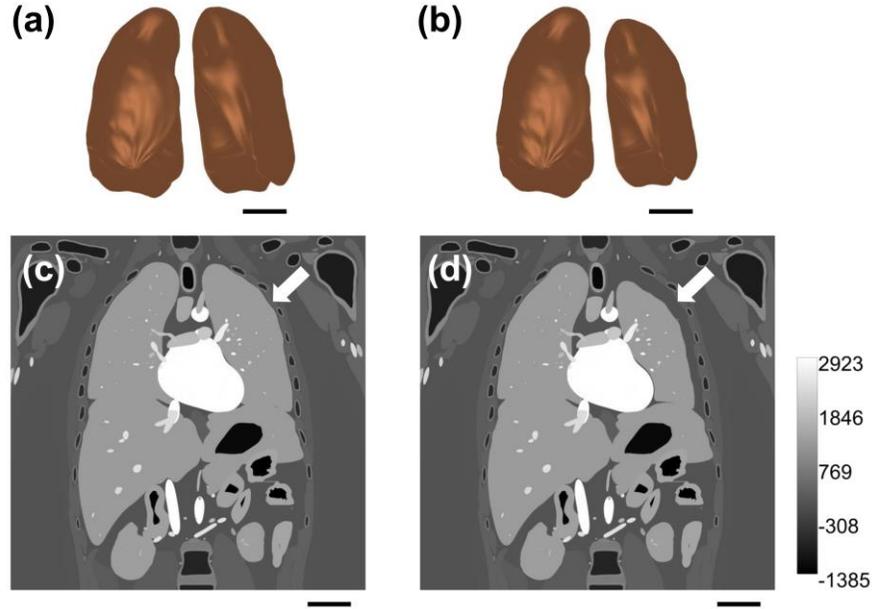

**Figure 4. Simulation of pulmonary atelectasis in the augmented XCAT phantom. (a) and (b) show the 3D rendered lung models before and after atelectasis introduced parenchymal volume loss in the left lung, respectively. (c) and (d) show examples of XCAT slices before and after introduction of atelectasis in the left lung, respectively. Scale bars in (a) through (d): 5 cm.**

Figures 4(a) and 4(b) show the 3D rendered views of the lungs before and after shrinking the left lung by 10%, respectively. Figures 4(c) and 4(d) show the XCAT slices after before and after the augmentation, respectively. To model interstitial edema in edematous lungs, the thickness of alveolus wall for the left lung was increased by 20% without changing the alveolar density. Figure 5(a) shows examples of the Voronoi grids for normal (i) and edematous (ii) lungs. For pneumonic lungs, 10% of the alveoli in the left lung were randomly selected and assigned the material properties for pus (similar to blood), without changing the alveolar density or the wall thickness. Figure 5(b) shows an example of the Voronoi grids for pneumonic lungs after the augmentation.

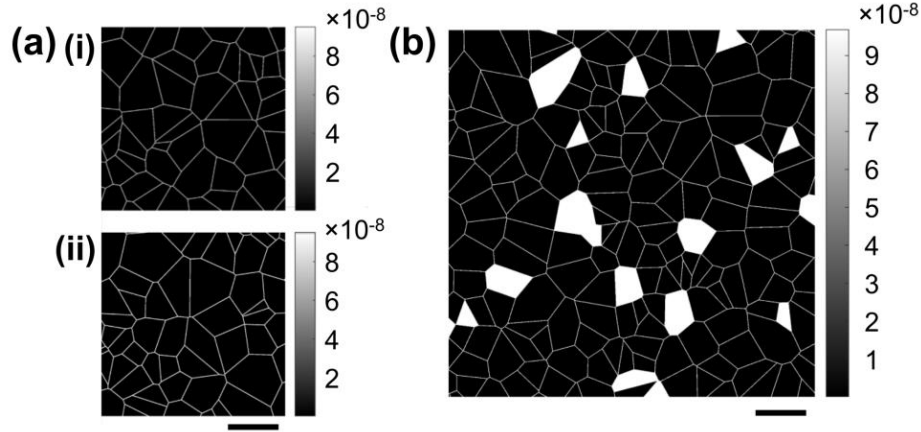

**Figure 5.** Examples of the lung microstructure simulated with Voronoi grids for (a) pulmonary edema and (b) patchy alveolar pneumonia. In (a), (i) is for normal lungs, which is shown together for comparison, and (ii) is for pulmonary edema with 20% increase in the interstitial wall thickness. In (b), 10% of randomly selected alveoli were assigned the material properties for blood (same as pus) to model alveolar pneumonia. Scale bars in (a) and (b): 1 mm.

**Table II. Material properties of the materials defined in the XCAT for the simulated X-ray energy of 50 keV.**

| Materials | Attenuation coefficient ($m^{-1}$) | Electron density ($\times 10^{26}$ $m^{-3}$) | Refractive index | |
|---|---|---|---|---|
| | | | $\delta$ ($\times 10^{-7}$) | $\beta$ ($\times 10^{-10}$) |
| Water | 22.7 | 3340 | 0.921 | 0.448 |
| Muscle | 23.7 | 3480 | 0.960 | 0.468 |
| Lung | 23.6 | 3450 | 0.951 | 0.466 |
| Dry spine | 48.3 | 4530 | 1.249 | 0.953 |
| Dry rib | 47.7 | 4500 | 1.241 | 0.940 |
| Adipose | 20.1 | 3180 | 0.877 | 0.397 |
| Blood | 24.2 | 3510 | 0.968 | 0.477 |
| Heart | 24.2 | 3510 | 0.968 | 0.477 |
| Kidney | 23.7 | 3480 | 0.960 | 0.468 |
| Liver | 24.1 | 3510 | 0.968 | 0.475 |
| Lymph | 23.5 | 3420 | 0.943 | 0.463 |
| Pancreas | 23.4 | 3460 | 0.954 | 0.462 |
| Intestine | 23.2 | 3420 | 0.943 | 0.457 |
| Skull | 61.0 | 5070 | 1.398 | 1.204 |
| Cartilage | 25.6 | 3620 | 0.998 | 0.506 |
| Brain | 23.6 | 3460 | 0.954 | 0.466 |
| Spleen | 24.1 | 3510 | 0.968 | 0.475 |
| Air | 0.025 | 3.622 | 0.000999 | 0.0004946 |
| Skin | 24.3 | 3600 | 0.993 | 0.480 |
| Red marrow | 22.7 | 3420 | 0.943 | 0.447 |
| Yellow marrow | 20.7 | 3280 | 0.905 | 0.408 |

| Testis  | 23.6 | 3460 | 0.954 | 0.466 |
| Thyroid | 24.9 | 3480 | 0.960 | 0.491 |
| Bladder | 23.7 | 3430 | 0.946 | 0.467 |

**Multi-layer wave-optics simulation of X-ray propagation through the augmented XCAT phantom and gratings**

Suppose that a planar X-ray wave of the energy 50 keV is propagating in the $z$ direction and incident onto the phantom perpendicular to the coronal plane (Figure 1). For the $N$-th layer, the transmittance function $T_N(x, y)$ is defined as follows.

$$T_N(x, y) = exp\{i(2\pi/\lambda_0)[n(x, y; N) - n_0]\Delta z\}, \qquad [3]$$

where $n(x, y; N)$ and $n_0$ are the complex refractive index of the $N$-th layer and the immersion medium (air), respectively, $\Delta z$ is the thickness of each layer, and $\lambda_0$ is the wavelength of the X-ray in the immersion medium ($2.4797 \times 10^{-11}$ m). Using axial averaging, $n(x, y; N)$ is averaged over multiple slices, and $\Delta z$ is increased. Assuming a plane wave, the wave field incident onto the first layer can be simply set as $U_0(x, y) = 1$. The wave field after the $N$-th layer can be calculated using the angular-spectrum scalar wave theory, Eq. (4).

$$U_N(x, y) = \mathcal{F}^{-1}\{H(u, v; \Delta z)\mathcal{F}\{U_{N-1}(x, y)T_N(x, y)\}\}, \qquad [4]$$

where $U_N(x, y)$ is the wave field after the $N$-th layer; $H(u, v; \Delta z)$ is the transfer function defined in Eq. (5); $\mathcal{F}$ and $\mathcal{F}^{-1}$ represent the 2D Fourier transform and the 2D inverse Fourier transform, respectively. The transfer function is given as follows.

$$H(u, v; \Delta z) = exp\{i2\pi\Delta z w(u, v)\}, \qquad [5]$$

where $\Delta z$ is the thickness of each layer, and $(u, v)$ are the spatial frequency components corresponding to $(x, y)$, respectively, and $w(u, v) = \sqrt{(n_0/\lambda_0)^2 - u^2 - v^2}$.

The wave field after the last layer of the phantom, $U^{(1)}(x, y)$, is incident onto the absorption grating $G_1$ of 10 μm period, which is in direct contact with the phantom (Figure 1). The wave field right after G1 can be written as

$$U^{(2)}(x,y) = U^{(1)}(x,y)[rect(x/p) * comb(x/p)], \qquad [6]$$

where $p$ is the grating period, and $rect(x) = 1$ (for $|x| < 1/2$); $1/2$ (for $|x| = 1/2$); and 0 otherwise. The comb function is defined as $comb(x) = \sum_{m=-\infty}^{\infty} \delta(x-m)$, where $\delta(x)$ is the Dirac delta function. The amplitude grating $G_2$ has the same period as $G_1$. Typically, $G_2$ is placed at a fractional Talbot distance from the grating $G_1$, which would correspond to 2.0164 m for the simulation condition used in this study. However, in the recent demonstrations of XDFI using a pig[38] and humans[39], $G_2$ was placed at a much shorter distance, which rendered the system to be much more compact yet different from a standard Talbot-Lau setup. In this simulation study, we place $G_2$ at the distance of $d = 0.25$ m from $G_1$, similar to these experimental demonstrations. The propagation of X-ray between $G_1$ and $G_2$ can be calculated using the angular-spectrum scalar wave theory[33]. The wave field right before $G_2$ can be written as

$$U_3(x,y) = \mathcal{F}^{-1}\{\tilde{U}^{(2)}(u,v)H(u,v;d)\}, \qquad [7]$$

where $\mathcal{F}^{-1}$ represents the 2D inverse Fourier transform, $\tilde{U}^{(2)}(u,v)$ is the 2D Fourier transform of $U^{(2)}(x,y)$, and $H(u,v;d) = exp\{i(2\pi/\lambda)d[1-(\lambda u)^2-(\lambda v)^2]^{1/2}\}$ is the transfer function for the light-field propagation over $d$. The intensity measured at the position is given by $I^{(3)}(x,y) = |U^{(3)}(x,y)|^2$.

To acquire the absorption, the differential phase, and the normalized visibility, 8 images are recorded while translating $G_2$ by different amounts. For the $k$-th position ($k = 1, 2, …, K$), the intensity after $G_2$ is given by

$$I^{(4)}(x,y;k) = I^{(3)}(x,y)[rect(x/p) * comb(x/p - k/K)]. \qquad [8]$$

The penumbral blurring due to the finite focal spot size of the X-ray source is modeled as an asymmetric Gaussian function, which is applied in the Fourier space. In the direction parallel to the grating lines, the width of the blurring function is determined by projecting the X-ray focal spot size (1.2 mm) onto the detector plane. In the direction orthogonal to the grating lines, the width of the blurring function is determined by projecting one period of the source grating $G_0$ (70 μm) onto the detector plane.

Noteworthy, to capture the small-angle X-ray diffraction from the lung alveoli, we perform the X-ray propagation from Eqs. (3) through (8) using an upsampled grid of 1.25 µm resolution, which is 80 times as high resolution as used for the XCAT simulation. Because the size of each upsampled image is too large, we divide the computation domain to 32×32 overlapping tiles, each 10880×10880 pixels. Then, to expedite the wave-optics simulation, we average every 20 slices along the beam propagation direction. The axial averaging converts 2505 slices to 126 layers, 2 mm thick each. For the 50 keV X-rays diffracted by a periodic structure of the 200 µm period, the diffraction angle is only 0.124 µrad. The X-ray image will be laterally shifted only by 0.248 nm after 2 mm, which is about 3 orders of magnitude smaller than the pixel resolution of our wave-optics simulation.

The detector, which comprises 1024×1024 pixels of 400 µm pixel size, is located right after $G_2$. Assuming a detector array of $N_d \times N_d$ pixels, each of which is a square with dimensions $\Delta \times \Delta$, and no dead space between pixels, the intensity recorded with the detector can be written at the $(i,j)$-th pixel ($i, j = 1, 2, …, N_d$) as

$$I^{(5)}[i,j;k] = \left(DI^{(4)}(x,y;k)\right)[i,j] := \int_{-\Delta/2}^{\Delta/2} \int_{-\Delta/2}^{\Delta/2} I^{(4)}(i\Delta + \xi, j\Delta + \eta; k) d\xi d\eta, \quad [9]$$

where $D$ represents an integration downsampling operator. The augmentation of lungs using Voronoi grids is the most computationally intensive step. Using parallel processing on 16 CPUs, with 24 GB memory for each, it took about 11 days on average to complete the computation. The computation for the fibrosis case, which involves increasing the alveolus thickness (i.e., Voronoi boundaries), took about 1.5 months. Currently, the augmentation using Voronoi grids is done on the fly, as it would require a lot of disk space to save all the augmented slices. One idea to overcome this challenge is to use a preselected set of Voronoi patterns (maybe 20–30), instead of using a completely random pattern for every slice. Figure 6 shows an example of 8 images recorded for different values of $k$.

**Extraction of absorption, differential phase, and dark-field images from the raw images for various positions of G$_2$**

From the 8 raw images, the absorption, the differential phase, and the dark-field images were calculated using the standard Fourier retrieval method[40]. For each pixel, a phase stepping curve was acquired, which is the intensity $I^{(5)}[i,j;k]$ as a function of the G$_2$ step position, $k = 1, 2, \ldots, 8$. These phase step curves sample the sinusoidal intensity pattern over one period of length $K$ (Eq. 8). For each pixel index, a 1D discrete Fourier transform was performed with respect to step index $k$. The X-ray intensity can be calculated from the zeroth order Fourier coefficient. The absorption due to the phantom, here presented as relative transmission $T$, can be calculated as the ratio of measured with the sample $a_{0,s}$ and the reference (i.e., without the sample) $a_{0,r}$.

$$T = \frac{a_{0,s}}{a_{0,r}}. \qquad [10]$$

The X-ray refraction in the direction perpendicular to the grating grooves can be quantified as the transverse shift of the interference fringes, which can be calculated from the phase of the first Fourier component, $\phi_1$. The X-ray refraction due to the phantom can be obtained from the relative phase shift between the sample and reference measurements.

$$\phi = \phi_{1,s} - \phi_{1,r} = \frac{\lambda d}{p} \frac{\partial \Phi}{\partial x}, \qquad [11]$$

where $x$ is the phase stepping direction. The ultra-small-angle scattering of X-rays can be quantified as the change in the visibility, which is given as the ratio of the first and the zeroth order Fourier coefficients.

$$V = a_1/a_0, \qquad [12]$$

The scramble of the wavefront due to the microstructure in the phantom can be obtained from the normalized visibility, the ratio of the visibility measured with the sample and the reference (i.e., without the sample).

$$V_{norm} = V_s/V_r. \qquad [13]$$

**Results and Discussions**

**Wave-optics simulation of XPCI and XDFI using the XCAT human phantom**

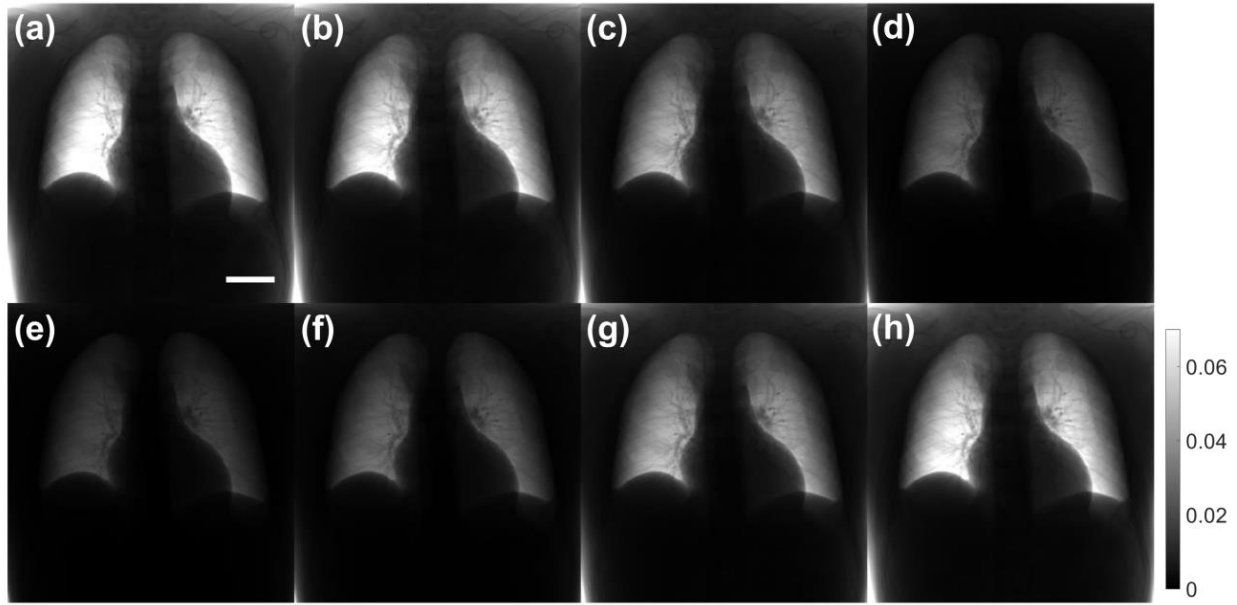

**Figure 6. GB-XPCI simulation of the augmented XCAT phantom: simulated intensity images for 8 different positions of $G_2$. The scale bar (5 cm) in (a) also applies to the images (b) through (h).**

Figure 6 shows an example of the raw images simulated for 8 different positions of $G_2$ or for different values of *k* representing different phases sampled from the Talbot pattern. Because the incident X-ray wave was assumed to have uniform amplitude of one, the values shown in Figs. 6(a) through 6(h) correspond to the intensity normalized with the background. As expected, the intensity values change with the grating position being the highest in Figure 6(a), where $G_1$ and $G_2$ are in phase, and the lowest in Figure 6(e), where $G_2$ is shifted with respect to $G_1$ by half the period. The amount of modulation in intensity also varies by the multiple scattering properties of the imaged tissue. In soft tissues of the mediastinum and abdomen there is a large modulation ranging from 0.45 in Figure 6(a) to 0.047 in Figure 6(e). However, due to the multiple-scattering interfaces of the lung, the modulation of the X-ray intensity in the lungs between Figure 6(a) and 6(e) is greatly reduced relative to other tissues. This is illustrated by the residual lung transmission in Fig 6(e) despite being absent in other tissues. Figure 7(a) shows the absorption (i), phase (ii), and dark-field (iii) images extracted from the 8 images shown in Figure 6 using the standard Fourier analysis (see Methods). The absorption image shows the absorbance, the negative logarithm of the transmittance with base 10, comparable to a conventional chest X-ray imaging. The phase image and the dark-field image are additional images provided by the grating-based XPCI and XDFI methods, respectively.

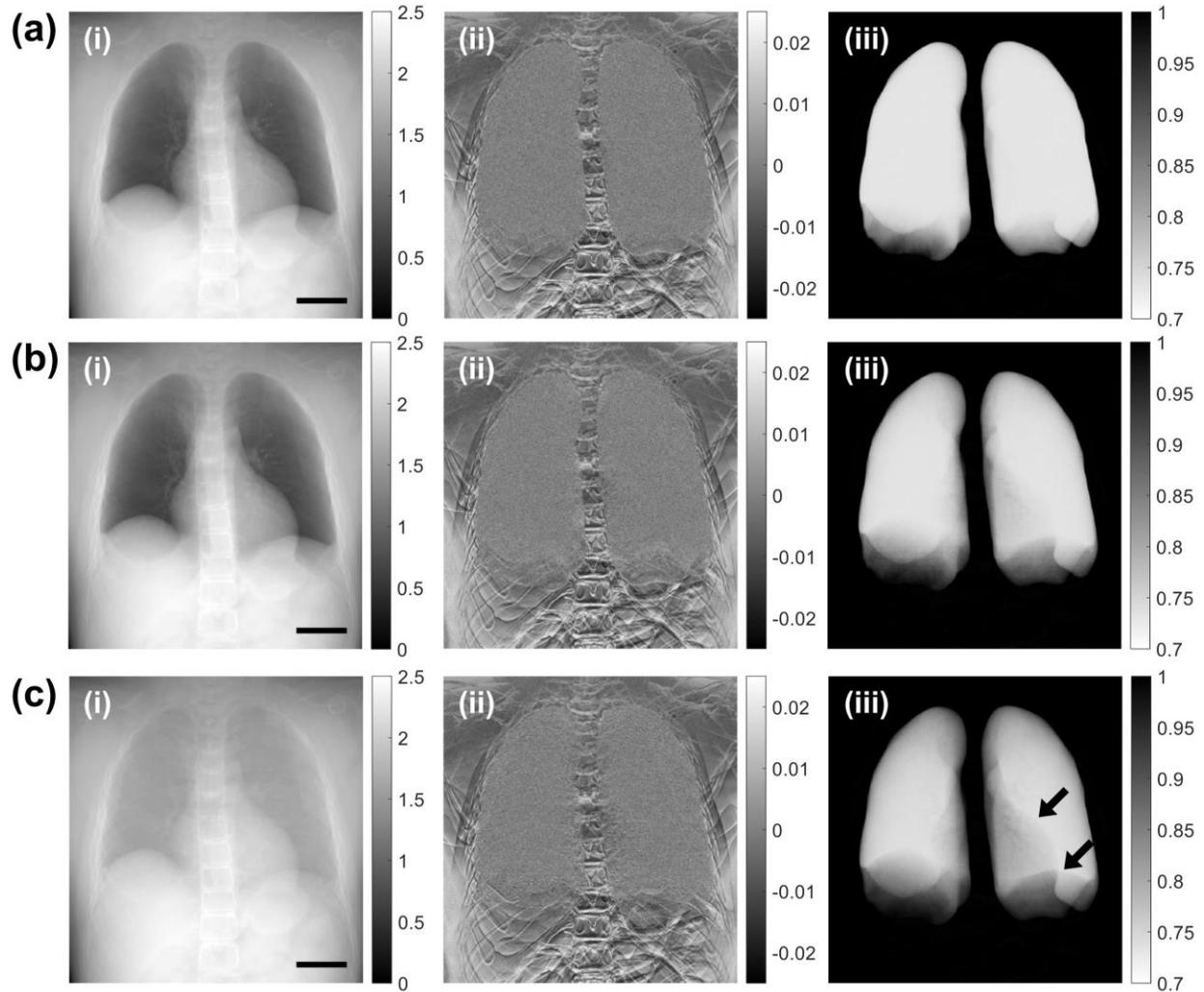

**Figure 7.** Absorption (i), differential phase contrast (ii), and dark-field (iii) images of the XCAT phantoms simulating normal lungs (a), emphysema (b), and pulmonary fibrosis (c), respectively. The scale bars (5 cm) in (i) also apply to (ii) and (iii).

**Dependence of absorption, phase, and dark-field signals on the microstructure of lungs**

Using the developed method, we have simulated the dark-field image of human lungs affected by fibrosis and emphysema. The diseased lungs were modeled by varying the cell number and edge thickness of Voronoi grids, as described in Methods. Figures 7(a) through 7(c) show the absorbance, phase, and dark-field images for normal, emphysematous, and fibrotic lungs. Figure 7(a) illustrates how the different signal contrasts available in grating interferometry vary by tissue type. The primary signal in absorption image (i) comes from the X-ray absorbing bones and soft tissues of the mediastinum and the thorax. The differential phase image (ii) shows strong gradient at tissue interfaces while the lungs appear as a high frequency noise due to the abundance of multiple-scattering interfaces that prevent reliable phase retrieval. However, this strong multiple scattering signal from the lung parenchyma provides abundant signal in the dark-field image (iii) with less anatomic background from the bones and heart shadow.

The effects of emphysema are illustrated in Figure 7(b). The reduced lung density in emphysema, modeled here as a lower density Voronoi grid, results in subtle reduction in absorption compared to the healthy example absorption image. Conversely, there is a greater visual difference in signal between the emphysema and healthy dark-field images compared to the subtle difference in absorption images. The fibrotic lungs, modeled as a Voronoi grid with lower density and thicker wall, show clearer changes in both the absorption and dark-field images, as shown in Figure 7(c). The enhanced sensitivity to structural changes in the lungs of X-ray dark-field is also supported by comparing the histograms of healthy and diseased lungs as well as the mean signals in the lungs in Figure 8.

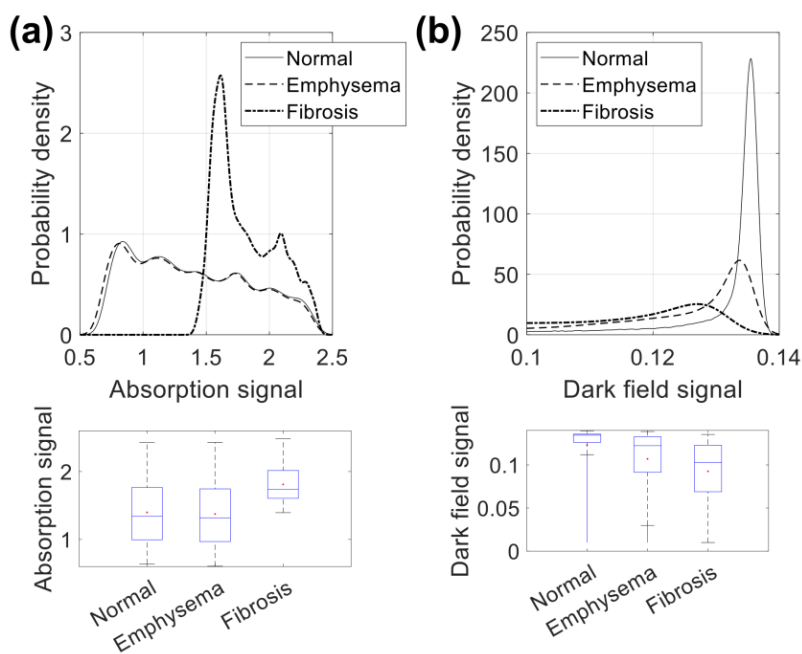

**Figure 8. Comparison of absorption (a) and dark-field (b) signals for normal, emphysematous, and fibrotic lungs. The probability density distributions and boxplots in the lung region are shown for the three cases. On each box plot, the central mark represents the median value, and the bottom and top edges represent the 25th and 75th percentiles, respectively.**

Figure 8(a) shows the probability distributions and boxplots of the attenuation signal (i.e., absorbance) within the lung region for the normal, emphysematous, and fibrotic cases. For the attenuation signal, the normal and emphysematous lungs produce wide distributions that are almost overlapping. The median values are also about the same: 1.342 and 1.319 for the normal and emphysematous lungs, respectively. The fibrotic lung produces a narrow distribution with a peak at 1.61, which is clearly distinguished from the other two cases, and a significantly higher median value of 1.741. Figure 8(b) shows the probability distributions and boxplots of the dark-field signal within the lung region for the same conditions. The normal, emphysematous, and fibrotic lungs show clearly distinct distributions on the histogram. The median absorbances for the three cases are 0.135, 0.122, and 0.103, respectively. Figure 9(a) shows the probability distributions of the dark-field signal within the lung region for emphysema at different severity levels: normal, mild, moderate, and severe.

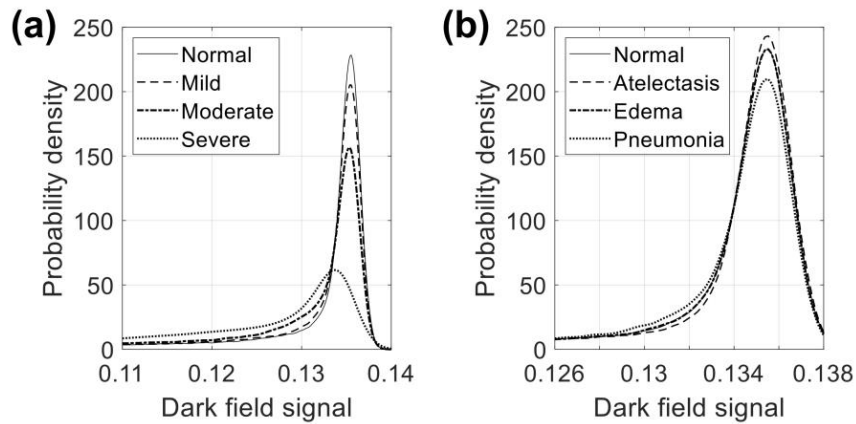

**Figure 9.** Comparison of the dark-field signals for emphysematous lungs at different levels of severity (a) and the dark-field signals for normal, atelectatic, edematous, and pneumonic lungs (b). The probability density distributions of the dark-field signal in the lung region are shown for the different cases.

We have also simulated atelectatic, edematous, and pneumonic lungs. Figure 9(b) shows the probability distribution of the dark-field signal within the lung region for these different disease states. This figure illustrates the limitation of using dark field signal alone in characterizing lung pathology. As can be seen, pathophysiologic alterations in lung structure induced by alveolar fluid accumulation in pneumonia, wall thickening in interstitial edema, and deflation in atelectasis all result in a reduction in dark-field signal. Therefore, using the dark-field image in isolation can be misleading. Signal alteration by emphysema and fibrosis is manifested by drop in dark-field signal relative healthy lung, as seen in Figure 8, and observation that agrees with experimental findings[41]. One should regard the absorption and dark field signals as complementary to each other, providing information about very different aspects of the lung. In order to enhance the sensitivity and specificity of X-ray imaging, researchers have proposed methods that combine both signals into a normalized scatter contrast for visualizing different lung pathologies[5].

The performance of grating interferometer X-ray dark-field and absorption contrasts at distinguishing different tissue properties is strongly dependent on the acquisition factors such as kV and phase stepping as well as grating geometry. While some studies have demonstrated optimizing grating interferometer setups experimentally[42], doing so is challenging and time-consuming as each geometry requires careful alignment, is dependent on the gratings available, and is application dependent. By incorporating a multi-layered wave-based approach to model multiple-scattering in X-ray dark-field imaging, the proposed framework enables system simulation and acquisition optimization of X-ray grating interferometers across a range of grating geometries and acquisition settings to optimize parameters for maximal performance in any given application.

While the study presented in this work featured a system geometry and set of acquisition parameters based on an experimental system,[39] once the numerical phantom is precomputed multiple system geometries and acquisitions can be rapidly investigated to find the optimal geometry and parameter set to maximize disease sensitivity and specificity. Availability of such modeling approaches can be used to design and optimize X-ray grating interferometers more rapidly for new and improved applications.

**Acknowledgments**

This work was funded by the National Institutes of Health (1R03EB032038). We gratefully acknowledge the UWM High Performance Computing facility for providing the computer time on the GPU cluster.

**Additional Information**

The authors declare no competing interest.